\documentstyle[11pt]{article}
\begin{document}
\begin{titlepage}
\title{ The Geometry of Self-Dual Gauge Fields}
\author{ {\bf A. H. Bilge}\\
{\small Department of Mathematics, Institute for Basic Sciences}\\
{\small TUBITAK Marmara Research Center}\\
{\small Gebze-Kocaeli, Turkey}\\
{\bf T. Dereli}\\
{\small Department of Physics}\\
{\small Middle East Technical University}\\{\small Ankara, Turkey}\\
{\bf \c{S}. Ko\c{c}ak}\\
{\small Department of Mathematics}\\
{\small Anadolu University}\\{\small Eski\c{s}ehir, Turkey}}
\maketitle
\begin{abstract}
Self-dual 2-forms in
 $D=2n$ dimensions are characterised by an eigenvalue criterion. The equivalence of various definitions of self-duality is proven.
We show that the self-dual 2-forms determine a $n^2-n+1$ dimensional manifold
${\cal{S}}_{2n}$ and the
dimension of the maximal linear subspaces of ${\cal{S}}_{2n}$ is equal to the
Radon-Hurwitz number of linearly independent vector fields on the sphere $S^{2n-1}$. The relation between the
maximal linear subspaces and the representations of
Clifford algebras is noted.
A general procedure based on this relation for the explicit construction of linearly self-dual 2-forms is given.
The construction of the octonionic instanton solution in $D=8$ dimensions
is discussed.
\end{abstract}
\end{titlepage}

\noindent {\bf 1. Self-duality as an eigenvalue criterion}
\vskip 2mm
Let $M$ be a $D=2n$ dimensional differentiable
manifold, and $E$ be a vector bundle over
$M$ with standard fiber $R^N$ and  structure group $G$.
The gauge potentials
can be represented by a $\cal G$-valued connection 1-form $A$ on $E$, where $\cal G$ is
a linear representation of the Lie algebra of  the gauge group $G$. Then the
gauge fields are represented by the
curvature of the connection $A$
that is given locally by the $\cal G$-valued 2-form
$$F=dA-A\wedge A .$$
The Yang-Mills action is the $L_2$ norm of the
curvature 2-form $F$
$$\| F\|^2=\int_M {\rm tr} (F \wedge *F)$$
where $*$ denotes the Hodge dual defined relative to a positive definite
metric on $M$.
The Yang-Mills equations
$$d_EF=0,\quad ^*d_E^*F=0,$$
where $d_E$ is the bundle exterior covariant derivative and
$-*d_E*$ is its formal adjoint, determine the critical points
of the action.

In $D=4$ dimensions
 $F$ is called self-dual or anti-self-dual provided
$$*F=\pm F.$$ In this case the
self-dual or anti-self-dual 2-forms are the global extrema of the Yang-Mills
action.  This is due to the fact that
the Yang-Mills action has a topological lower bound:
$$\|F\|^2 \geq \int_M {\rm tr} (F\wedge F).$$
The term ${\rm tr} (F\wedge F)$ is related to the Chern classes
of the bundle. Actually if $E$ is a
complex 2-plane bundle with $ c_1(E)=0$, then
the topological bound is proportional to $ c_2(E)$ and
this lower bound is realised by a (anti-)self-dual
connection.
Furthermore, $SU(2)$ bundles over a four manifold are classified by $\int
c_2(E)$, hence self-dual connections are minimal representatives of the
connections in each  equivalence class of $SU(2)$ bundles.
This is a generalisation of the fact that an $SU(2)$ bundle admits a flat
connection if and only if it is trivial.

In order to derive topological bounds in higher dimensions
we briefly recall the computation of the characteristic classes of a vector
bundle $E$ in terms of the local curvature 2-forms [1].
Let  $F^\alpha$ be the matrix of the local curvature 2-form  with respect to a
local basis of sections of $E$  on
a trivialising neigbourhood $U_\alpha$.
 The invariant polynomials $\sigma^\alpha_k$ of $F^\alpha$ are defined by
$${\rm det}(I+tF^\alpha)=\sum_{k=0}^n\sigma^\alpha_k t^k.$$
Hence they are independent of the basis of local sections. Thus if
$\sigma^\alpha_k$ and $\sigma_k^\beta$ are invariant polynomials of
the local curvature 2-forms $F^\alpha$ and $F^\beta$, on
$U_\alpha$ and $U_\beta$,respectively, then  they agree on the intersection $U_\alpha\cap
U_\beta$. Hence these  locally defined $2k$-forms patch up to give globally
defined
$2k$-forms $\sigma_k$. Furthermore it can be shown that
the $\sigma_k$ 's are closed $2k$-forms.
They realise  the
de Rham cohomology classes in $H^{2k}$. These
cohomology
classes depend only on the bundle, i.e. they are independent of the connection.
For a complex vector bundle, the cohomology class of $\sigma_k$  is
proportional to the Chern class $c_k$, while for a real vector bundle, the
$\sigma_{2k+1}$'s  are exact forms, and $\sigma_{2k}$'s are proportional to the
Pontrjagin classes $p_k$'s. Furthermore, for an
$SO(N)$ bundle, the
square root of the determinant of $F$ (which is a ring element) defines the
Euler class $\chi$.
In order to avoid proportionality constants, we will
work with the quantities $\sigma_k$'s instead of the Chern or Pontrjagin
classes.
 The $\sigma_k$'s can also be written as
linear combinations of $trF^k$ (see for example [2] Vol.1, p.87), where $F^k$
means the product of the
matrix $F$ with itself $k$ times, with the wedge multiplication of the entries.
In the following we consider real $SO(N)$ bundles.

In $D=4$ dimensions the topological bound we wrote above is the only one that is available. On the other hand in $D=8$ dimensions
it is possible to introduce two independent topological bounds.
The topological lower bound on the action
$$\int_{M} tr(F^2 \wedge ^* F^2)  \ge  \int_M p_{2}(E)$$
is well-known. The
self-duality  of $F^2$ in the Hodge sense gives global minima of this
action involving the second Pontryagin number    $\int p_2(E)$.
We introduced [3] another
topological lower bound on the action
$$\int tr(F\wedge ^* F)^2\ge \frac{2}{3} \int_{M}  p_1(E)^2.$$
This involves the square of the first Pontryagin number and has to be taken into account as the topology of the Yang-Mills bundle
on an eight manifold has to be characterised by both the first and the second  Pontryagin numbers.
In general, similar topological bounds could be given in higher dimensions.
We now turn to the definition of self-dual gauge fields  in higher dimensions [3],[4] that would saturate
these topological bounds.

\proclaim Definition 1. Suppose $\omega$ is a real 2-form in $D=2n$ dimensions,
and let $\Omega$ be the corresponding $2n \times 2n$
skew-symmetric matrix with respect to some local orthonormal basis.
Let
$\pm i\lambda_{1},...,\pm i\lambda_{n}$ be the eigenvalues of $\Omega$.
$\omega$ is said to be (anti-)self-dual if
$$|\lambda_{1}| = |\lambda_{2}| = \dots  = |\lambda_{n}|.$$

\noindent There are $2^n$ possible ways to satisfy the above set
of equalities. The half of these correspond to self-dual
2-forms,  while  the  remaining  half  correspond  to anti-self-dual
2-forms.

It is not difficult to check that in $D=4$ dimensions, the above definition
coincides with the usual definition of self-duality in the Hodge sense.
Let $\Omega$ be the skew symmetric
matrix representing a 2-form $F$ in four dimensions. Then it can be seen that
the eigenvalues of the matrix $\Omega$ satisfy
$$
\lambda_1\mp\lambda_2=\sqrt{(\Omega_{12}\mp \Omega_{34})^2+
(\Omega_{13}\pm \Omega_{24})^2
         +(\Omega_{14}\mp \Omega_{23})^2}
$$
Thus for self-duality
$\lambda_1= \lambda_2$,
while for anti-self-duality
$\lambda_1= -\lambda_2$.
In both cases the absolute values of the
eigenvalues are equal. Two cases are distinguished by the sign of the Pfaffian
of $\Omega$:
$$
\Omega_{12}\Omega_{34}-\Omega_{13}\Omega_{24}+
\Omega_{14}\Omega_{23}.
$$

There were several attempts
to generalize the notion of self-duality to higher dimensions:

i) A 2-form $\omega$ in  $D=2n$ dimensions
is called self-dual if the Hodge dual of $\omega$ is proportional to
$\omega^{n-1}$.  Here  wedge  product of $\omega$'s should be understood.
This notion is introduced by Trautman [5], and
Thcrakian [6] and used widely by others.

ii) A self-dual 2-form $\omega$ in $D=4k$ dimensions
is defined to be the one such that $\omega^{k}$ is self-dual in the
Hodge sense,
that is $^*\omega^k = \pm \omega^k$.
This notion is also introduced by Thcrakian [6] and adopted by
Grossman, Kephardt and Stasheff (GKS) in their study of
octonionic instantons in eight dimensions [7].

iii) Both the criteria above  are non-linear.
Alternatively, (anti-)self-dual 2-forms in $D=2n$ dimensions
can be defined as eigen-bivectors of a completely antisymmetric fourth rank tensor that is invariant under a subgroup of $SO(2n)$. The
set of such self-dual 2-forms
would span a linear space. This notion of self-duality is introduced
 by Corrigan, Devchand, Fairlie and Nuyts (CDFN )
who studied  the first-order  equations satisfied by  Yang-Mills  fields
in spaces of dimension
greater than four  and derived
$SO(7)$ self-duality equations in $R^8$ [8].

It can be shown that any self-dual 2-form defined by the above criteria
satisfies the  Yang-Mills equations.
However, the corresponding   Yang-Mills action
need not be extremal.
For details we refer to review articles [9],[10].
Here we will prove that our eigenvalue criterion
encompasses all these three notions of self-duality.

We start by noting that the invariant
polynomials $s_{2k}$  of $\omega$ can be expressed in terms of the elementary
symmetric functions of the $(\lambda_k)^2$'s.
Then the inner products
$(\omega^k,\omega^k)$ and  $s_{2k}$'s are related as follows:
\begin{eqnarray}
(\omega,\omega)&=&s_2=\lambda_1^2+\lambda_2^2+\dots
+\lambda_n^2,\cr
\frac{1}{(2!)^2}(\omega^2,\omega^2)&=&
s_4=\lambda_1^2\lambda_2^2+\lambda_1^2\lambda_3^2+\dots
                +\lambda_{n-1}^2\lambda_n^2, \cr
\frac{1}{ (3!)^2}
(\omega^3,\omega^3)&=& s_6=\lambda_1^2\lambda_2^2\lambda_3^2
                +\lambda_1^2\lambda_2^2 \lambda_4^2+\dots
                +\lambda_{n-2}^2\lambda_{n-1}^2 \lambda_n^2,\cr
&.&\cr &.& \cr &.& \cr
\frac{1}{ (n!)^2}(\omega^n,\omega^n)=
\frac{1}{(n!)^2}\mid*\omega^{n}\mid^2&=& s_{2n}=
               \lambda_1^2\lambda_2^2\dots       \lambda_n^2. \nonumber
\end{eqnarray}
If we define the weighted elementary
symmetric polynomials by
$${n\choose k}q_k= s_{2k},$$
we have the inequalities
(see for example [11], Ch.2, Sec. 3).
$$q_1\ge q_2^{1/2}\ge q_3^
{1/3}\ge \dots \ge q_n^{1/n},\quad\quad
q_{r-1}q_{r+1}\le q_r^2, \quad 1\le r<n,$$
and the equalities hold iff all the $\lambda_k$'s are equal, hence in the case
of self-duality. We have the following lemma.

\proclaim Lemma 2. Let $\omega$ be a 2-form in $2n$ dimensions. Then
$$(n-1)
(\omega,\omega)^2-{n\over 2}(\omega^2,\omega^2)\ge0$$
and $$(\omega^{n/2},\omega^{n/2})\ge *\omega^n,$$
provided $n$ is even. Equality holds if and only if all eigenvalues of $\omega$ are equal.

\noindent {\it Proof.} To obtain the first inequality we use,
\begin{eqnarray}
q_1^2&\ge &q_2\cr
{1\over n^2} s_2^2&\ge&{2\over n(n-1)} s_4\cr
{1\over n^2}(\omega,\omega)^2&\ge&{2\over n(n-1)}{1\over
4}(\omega^2,\omega^2)\cr \nonumber
\end{eqnarray}
which gives the desired result. Similarly using
\begin{eqnarray}
q_{n/2}^2&\ge & q_n\cr
\left({(n/2)!(n/2)!\over n!} s_{n/2}\right)^2&\ge&s_{2n}\cr
\left({1\over n!}(\omega^{n/2},\omega^{n/2})\right)^2&\ge&
{1\over (n!)^2}\mid *\omega^n\mid^2 \nonumber
\end{eqnarray}
we obtain the second inequality.\quad \quad e.o.p.

From Lemma 2,  we immediately have
\proclaim Corollary 3. A 2-form $\omega$ is  self-dual iff
$\omega^{n/2}$ is self-dual in the Hodge sense.

\proclaim Proposition 4. Let $\omega $ be a 2-form in $2n$ dimensions.
$$\omega^{n-1}=\kappa *\omega$$ where $\kappa$ is a constant, iff
$\omega$ is  self-dual and
$\kappa={n!\over n^{n/2}}
(\omega,\omega)^{{n\over 2}-1}$.

\noindent {\it Proof.}

If $\omega$ is self-dual, we can choose an orthonormal basis such that
$\omega=e_1e_2+e_3e_4+e_5e_6+e_7e_8$ with respect to this basis, and it can
easily be seen that the identity holds.
Conversely, if the identity holds, then
multiplying it with $\omega$ and taking Hodge duals,
we obtain,
$*\omega^n=\kappa(\omega,\omega)$. Since $(\omega,\omega)= s_2=nq_1$ and
$\mid*\omega^n\mid=n! s_{2n}^{1/2}=n!q_n$, we obtain $\kappa=(n-1)!q_n^{1/2}/q_1$.
Then
taking inner products
of both sides of the identity with themselves, we obtain
$(\omega^{n-1},\omega^{n-1})=k^2(*\omega,*\omega)=k^2(\omega,\omega)$.
Substituting the value of $\kappa$ obtained above,  and using
$(\omega^{n-1},\omega^{n-1})=\big((n-1)!\big)^2nq_{n-1}$, we obtain
$q_n= q_{n-1}q_1$. But since $q_1\ge q_n^{1/n}$, we have $q_n\ge
q_{n-1}q_n^{1/n}$, which leads to $q_n^{n-1}\ge q_{n-1}^n$. This is just
the reverse of the inequality proved in the previous lemma
, hence equality must hold and furthermore
all eigenvalues of $\omega$ are equal in absolute value.
Thus  $\omega$ is  self-dual and
it can also  be seen that
$\kappa={n!\over n^{n/2}}
(\omega,\omega)^{{n\over 2}-1}$. \quad e.o.p.

\noindent We remark  that if $(\omega,\omega) $ is nonzero everywhere on $M$,
then it can be normalised to
have constant norm and it defines an almost complex structure.
In this case $*\omega=\kappa \omega^{n-1}$, where $\kappa$ is constant.
Consequently, if
$\omega$ is closed and has constant norm, then $*\omega$ is harmonic.
\vskip 3mm
\noindent {\bf 2. The explicit construction of self-dual 2-forms}${}^{[4]}$
\vskip 2mm
Let ${\cal S}_{2n}$ be the set of self-dual 2-forms in $2n$ dimensions.
If ${\bf A}_{2n}$ denotes the set of antisymmetric matrices in
$2n$ dimensions, then ${\cal S}_{2n}=\{ A\in {\bf A}_{2n}
\mid A^2+\lambda^2I=0,\lambda\in {\bf R}, \lambda \ne 0\}$.
Here and in the following $I$ denotes
an identity
matrix of appropriate dimension.

\proclaim Proposition 5. ${\cal S}_{2n}$ is diffeomorphic to the homogeneous
manifold $\big(O(2n)\times {\bf R}^+\big)/U(n) \times \{ 1 \}$,
and {\it dim}${\cal S}_{2n} =n^2-n+1$.

\noindent As $O(2n)$ has two connected components ( $SO(2n)$ and $O(2n) \setminus
SO(2n)$), $U(n)$ is connected and $U(n) \subset SO(2n)$,
${\cal S}_{2n}$ has two connected components. One of which
consists of the self-dual forms and the other of the
anti-self-dual forms.

Let ${\cal L}_{2n}^\alpha$ be a maximal linear subspace  of ${\cal S}_{2n}$,
where
$\alpha$ is a real parameter. The elements of ${\cal L}_{2n}^\alpha $
are
skew-symmetric and non-degenerate. Therefore,
\proclaim Proposition 6. The dimension of the maximal linear subspaces of
${\cal S}_{2n}$ is  equal to the number of linearly independent
vector fields on $S^{2n-1}$.

\noindent The maximal number  of pointwise linearly independent vector
fields on the sphere $S^N$ is given by the Radon-Hurwitz number $k$. If
$N+1=2n=(2a+1)2^{4d+c}$ with $c=0,1,2,3$, then $k=8d+2^c-1$.
Using this formula it can be seen that there are
three vector fields on $S^3$, seven on $S^7$, three on $S^{11}$,
eight on $S^{15}$ and so on. In particular
there is only one vector field on the spheres $S^{2n-1}$ for odd n.
This property shows that there is an intimate relationship between
generalised self-duality and Clifford algebras [12].

We shall now discuss a general procedure for constructing  linear subspaces of
self-dual forms.
Note that ${\cal S}_{2n}$ is the set of  skew-symmetric matrices
in $O(2n)\times R$. We define ${\cal P}_{2n}$ to be the set of symmetric
matrices in $O(2n)\times R$. Recall that an orthonormal basis for a
$k$-dimensional linear subspaces of ${\cal S}_{2n}$ corresponds to
the representation of $Cl_k$ in the skew-symmetric matrices. Similarly an
orthonormal  basis
for a $k$-dimensional linear subspace of ${\cal P}_{2n}$ corresponds to a
representation of the dual  Clifford algebra  $Cl'_k$ in the symmetric
matrices. These
bases will be the building blocks for self-dual forms in the double dimension.

We have already shown that in dimensions $2n=2(2a+1)$ the maximal  linear
subspaces of ${\cal S}_{2n}$ are one dimensional. Similarly, in
dimensions
$2n=4(2a+1)$, the dimension of maximal linear subspaces of ${\cal S}_{2n}$ are
three dimensional. It can be seen that the matrices
$$J_0=\pmatrix{ 0& 0& I& 0\cr
                0& 0& 0& I\cr
               -I& 0& 0& 0\cr
                0&-I& 0& 0\cr},\quad
  J_1=\pmatrix{ 0& I& 0& 0\cr
               -I& 0& 0& 0\cr
                0& 0& 0&-I\cr
                0& 0& I& 0\cr},\quad
  J_2=\pmatrix{ 0& 0& 0& I\cr
                0& 0&-I& 0\cr
                0& I& 0& 0\cr
               -I& 0& 0& 0\cr},\quad$$
where $I$ is the identity matrix, form an orthonormal basis
for three dimensional linear subspaces of ${\cal S}_{4(2a+1)}$.

Now we consider the only remaining case of self-dual 2-forms in
$8n$ dimensions.
The matrix of a self-dual form can be written in the form
$$\Omega =\pmatrix{A_a&B_a+B_s\cr
             B_a-B_s&D_a\cr},$$
where the matrices $A_a$, $B_a$, $D_a$'s are anti-symmetric and  $B_s$ is
symmetric. The requirement that $\Omega^2$ be proportional to
the identity matrix gives the following equations:
$$           A_a^2= D_a^2,\quad\quad A_a^2+B_a^2-B_s^2=kI,\quad\quad
               [B_a,B_s]=0,$$
$$           A_aB_a +B_aD_a=0,\quad \quad B_aA_a+D_aB_a=0,$$
$$           A_aB_s +B_sD_a=0,\quad \quad B_sA_a+D_aB_s=0.$$
If we furthermore require that $\Omega$ be build up from the linear
subspaces of ${\cal S}_{4n}$ and ${\cal P}_{4n}$, then we see that $A_a$,
$D_a$, $B_a$, $B_s$ have to be nondegenerate.

We shall give now an explicit construction of various linear subspaces of
${\cal S}_8$. Let ${\cal A}^-$ and ${\cal A}^+$ be orthonormal bases for linear
subspaces of ${\cal S}_{2n}$ and ${\cal P}_{2n}$, respectively.

In two dimensions we have the following structure.
$${\cal A}^-=\left\{\pmatrix{0&1\cr-1&0\cr}\right\},\quad
  {\cal A}_{(1)}^+=\left\{\pmatrix{1&0\cr0&1\cr}\right\},\quad
  {\cal
A}_{(2)}^+=\left\{\pmatrix{1&0\cr0&-1\cr},\pmatrix{0&1\cr1&0\cr}\right\}.$$

\noindent  From the commutation relations it can be seen that the orthonormal
bases for linear subspaces of self-dual 2-forms in four dimensions are
determined by the choice of $B_s$. The choice $B_s\in{\cal A}_{(1)}^+$ leads
to the usual anti self-dual 2-forms, while the choice $B_s\in{\cal A}_{(2)}^+$
leads to the self-dual 2-forms.  Hence in four dimensions we obtain two
different sets of orthonormal bases for linear subspaces of ${\cal S}_4$.
By similar considerations, we obtain seven different bases for linear
subspaces of ${\cal P}_4$. The elements of these bases are listed below:

$$
a_1=\pmatrix{ 0& 1 & 0 & 0\cr
             -1& 0 & 0 & 0\cr
              0& 0 & 0 & 1\cr
              0& 0 &-1 & 0\cr},\quad
a_2=\pmatrix{ 0& 0 & 1 & 0\cr
              0& 0 & 0 &-1\cr
             -1& 0 & 0 & 0\cr
              0& 1 & 0 & 0\cr},\quad
a_3=\pmatrix{ 0& 0 & 0 & 1\cr
              0& 0 & 1 & 0\cr
              0&-1 & 0 & 0\cr
             -1& 0 & 0 & 0\cr},\quad$$
$$b_1=\pmatrix{ 0& 1 & 0 & 0\cr
             -1& 0 & 0 & 0\cr
              0& 0 & 0 &-1\cr
              0& 0 & 1 & 0\cr},\quad
b_2=\pmatrix{ 0& 0 & 1 & 0\cr
              0& 0 & 0 & 1\cr
             -1& 0 & 0 & 0\cr
              0&-1 & 0 & 0\cr},\quad
b_3=\pmatrix{ 0& 0 & 0 & 1\cr
              0& 0 &-1 & 0\cr
              0& 1 & 0 & 0\cr
             -1& 0 & 0 & 0\cr},\quad$$
$$c_1=\pmatrix{ 0& 0 & 0 & 1\cr
              0& 0 &-1 & 0\cr
              0&-1 & 0 & 0\cr
              1& 0 & 0 & 0\cr},\quad
c_2=\pmatrix{ 0& 0 & 1 & 0\cr
              0& 0 & 0 & 1\cr
              1& 0 & 0 & 0\cr
              0& 1 & 0 & 0\cr},\quad
p_1=\pmatrix{ 0& 1 & 0 & 0\cr
              1& 0 & 0 & 0\cr
              0& 0 & 0 & 1\cr
              0& 0 & 1 & 0\cr},\quad$$
$$p_2=\pmatrix{ 1& 0 & 0 & 0\cr
              0&-1 & 0 & 0\cr
              0& 0 & 1 & 0\cr
              0& 0 & 0 &-1\cr},\quad
d_1=\pmatrix{ 0& 0 & 0 & 1\cr
              0& 0 & 1 & 0\cr
              0& 1 & 0 & 0\cr
              1& 0 & 0 & 0\cr},\quad
d_2=\pmatrix{ 0& 0 & 1 & 0\cr
              0& 0 & 0 &-1\cr
              1& 0 & 0 & 0\cr
              0&-1 & 0 & 0\cr},\quad$$
$$q_1=\pmatrix{ 0& 1 & 0 & 0\cr
              1& 0 & 0 & 0\cr
              0& 0 & 0 &-1\cr
              0& 0 &-1 & 0\cr},\quad
q_2=\pmatrix{ 1& 0 & 0 & 0\cr
              0&-1 & 0 & 0\cr
              0& 0 &-1 & 0\cr
              0& 0 & 0 & 1\cr},\quad
e_1=\pmatrix{ 1& 0 & 0 & 0\cr
              0& 1 & 0 & 0\cr
              0& 0 &-1 & 0\cr
              0& 0 & 0 &-1\cr}.\quad$$

Using the commutation relations it can be shown that in four  dimensions we
have the following orthonormal bases for the linear subspaces of ${\cal
S}_4$.
$${\cal A}_{(1)}^-=\{a_1,a_2,a_3\},\quad\quad
{\cal A}^-_{(2)}=\{b_1,b_2,b_3\},$$
$${\cal A}^+_{(1)}=\{I\},$$
$${\cal A}_{(2)}^+=\{c_1,c_2,e_1\},\quad\quad
  {\cal A}_{(3)}^+=\{p_1,q_2,d_2\},\quad\quad
  {\cal A}_{(4)}^+=\{p_2,q_1,d_1\},\quad\quad $$
$${\cal A}_{(5)}^+=\{c_1,p_1,p_2\},\quad\quad
  {\cal A}_{(6)}^+=\{c_2,q_2,q_1\},\quad\quad
  {\cal A}_{(7)}^+=\{e_1,d_2,d_1\},\quad\quad $$

Orthonormal bases for linear subspaces of ${\cal S}_8$ can be constructed using
the sets given above.
We now show that the basis obtained by choosing $B_s=I$ corresponds to the
representation of ${\cal C}l_7$ using octonionic multiplication.
 Let us describe an octonion by a pair of
quaternions $(a,b)$. Then the octonionic multiplication  rule is
$(a,b) \circ (c,d)=(ac-\bar{d}b,da+b\bar{c})$. If we represent an octonion
$(c,d)$ by a vector in $R^8$, its multiplication by imaginary octonions
correspond to linear transformations on $R^8$. Using the multiplication rule
above, it is easy to see that we have the following correspondences:

$$(i,0)\rightarrow\pmatrix{b_1&0\cr0&-b_1\cr},\quad\quad
  (j,0)\rightarrow\pmatrix{b_2&0\cr0&-b_2\cr},\quad\quad$$
$$(k,0)\rightarrow\pmatrix{b_3&0\cr0&-b_3\cr},\quad\quad
  (0,1)\rightarrow\pmatrix{  0&  I\cr  -I&0\cr}\equiv J,\quad\quad$$
$$  (0,i)\rightarrow\pmatrix{  0&a_1\cr a_1&0\cr},\quad\quad
  (0,j)\rightarrow\pmatrix{  0&a_2\cr a_2&0\cr},\quad\quad
  (0,k)\rightarrow\pmatrix{  0&a_3\cr a_3&0\cr}.\quad\quad$$

\noindent Thus we obtain in $D=8$ dimensions the following
matrix of a self-dual 2-form:
$$ \Omega = F_{12} J + \pmatrix{ \Omega'& \Omega''\cr {\Omega''} & -\Omega'\cr}$$
where $\Omega'$ is a D=4 self-dual 2-form , $\Omega''$ is a D=4
anti-self-dual 2-form and $F_{12}$ is a real function.

On the other hand a CDFN self-dual 2-form $F$ is obtained by imposing
the following  conditions among its components [8]:
\begin{eqnarray}
&F_{12}-F_{34}=0\quad F_{12}-F_{56}=0\quad F_{12}-F_{78}=0\cr
&F_{13}+F_{24}=0\quad F_{13}-F_{57}=0\quad F_{13}+F_{68}=0\cr
&F_{14}-F_{23}=0\quad F_{14}+F_{67}=0\quad F_{14}+F_{58}=0\cr
&F_{15}+F_{26}=0\quad F_{15}+F_{37}=0\quad F_{15}-F_{48}=0\cr
&F_{16}-F_{25}=0\quad F_{16}-F_{38}=0\quad F_{16}-F_{47}=0\cr
&F_{17}+F_{28}=0\quad F_{17}-F_{35}=0\quad F_{17}+F_{46}=0\cr
&F_{18}-F_{27}=0\quad F_{18}+F_{36}=0\quad F_{18}+F_{45}=0\cr \nonumber
\end{eqnarray}
\noindent We will  refer to the plane consisting of these forms as the
CDFN-plane.  The skew-symmetric matrix $\Omega$ of such a self-dual 2-form is
$$ \left ( \begin{array}{cccccccc}
0& \Omega_{12}&\Omega_{13}&\Omega_{14}&\Omega_{15}&\Omega_{16}&
\Omega_{17}&\Omega_{18}\\
 &  0&\Omega_{14}& -\Omega_{13}&\Omega_{16 }&-\Omega_{15}&\Omega_{18}&
-\Omega_{17}\\
 & &  0 &\Omega_{12}& \Omega_{17}&-\Omega_{18}&-\Omega_{15}&\Omega_{16}\\
 & & & 0& -\Omega_{18}&-\Omega_{17}&\Omega_{16}&\Omega_{15}\\
& & & & 0&\Omega_{12}&\Omega_{13}&-\Omega_{14}\\
 & & & & & 0&-\Omega_{14}&-\Omega_{13}\\
 & & & & & & 0 & \Omega_{12}\\
 & & & & & & &  0 \end{array} \right ). $$
It is easy to show that the above matrix is related to
our self-dual 2-form  by conjugation $ R^t \Omega R$ with
$$R= \left ( \begin{array}{cccccccc}
1&0&0&0&0&0&0&0\\
0&0&1&0&0&0&0&0\\
0&0&0&0&1&0&0&0\\
0&0&0&0&0&0&1&0\\
0&1&0&0&0&0&0&0\\
0&0&0&1&0&0&0&0\\
0&0&0&0&0&1&0&0\\
0&0&0&0&0&0&0&1 \end{array} \right ). $$

\noindent In fact, the choice $B_s\in \{d_2,p_1,q_2\}$ determines the
possible choices for $B_a$'s, $A_a$'s and $D_a$'s and leads
to the CDFN plane [4].

Finally, we would like to point out that
these constructions can be generalised
to dimensions which are multiples of eight, by replacing the unit element with
identity matrices of appropriate size.
In dimensions which are multiples of $16$, one can make use of the
property ${\cal C}l_{k+8}={\cal C}l_k\otimes {\cal C}l_8$ to obtain  a ${\cal
C}l_{k+8}$ representation on $R^{16n}$, using  an already known representation
of ${\cal C}l_k$ on $R^n$.
Hence linear subspaces of ${\cal S}_{16n}$ can be obtained
from the knowledge of
the linear subspaces of ${\cal S}_n$.
\vskip 6mm
\noindent {\bf 3. The octonionic instanton solution in $D=8$ dimensions}
\vskip 2mm

In this section we shall discuss the construction of the
octonionic instanton solution  [7],[13-17].
Before that let us give identities concerning
self-dual 2-forms.
We have already shown that a self-dual 2-form $\omega$ satisfies the
basic equalities:
$$\textstyle(\omega,\omega)^2= {2\over 3} (\omega^2,\omega^2)= {2\over
3}*\omega^4,$$
and
$$\textstyle\omega^3={3\over 2} (\omega,\omega)*\omega.$$
We can obtain from these
a series of identities concerning the product of two self-dual 2-forms.
We give in what follows some essential results without proof [18].

\proclaim Lemma 7.
Let $\omega$ and $\eta$ be  self-dual 2-forms. If
$(\omega,\eta)=0$ then
$$\omega^3\eta=0.$$
If furthermore $\omega \pm \eta$ are also self-dual then
$$(\omega,\omega)(\eta,\eta)=2(\omega^2,\eta^2)=2\omega^2\eta^2,$$
$$\omega^2\eta=\textstyle \frac{1}{2} (\omega,\omega)*\eta,$$
and
$$\omega\eta=*(\omega\eta).$$

\noindent Further applying the above equalities to three mutually orthogonal
(linear) self-dual 2-forms
we obtain

\proclaim Lemma 8. Let $\omega$, $\eta$ and $\alpha $ be mutually
orthogonal self-dual 2-forms such that $\omega+\eta+\alpha$ is also self-dual.
Then
$$\omega \eta \alpha=0.$$

\noindent Collecting these results we state the following

\proclaim Proposition 9.
Let $F=\sum \omega_a E_a$ where $\{\omega_a \}$ is a set of mutually orthogonal
(linear) self-dual 2-forms and $\{E_a\}$ is
a basis of the Lie algebra of a gauge group $G$. Then \\
\noindent i)  $F^2=*F^2$ for any $G$,\\
\noindent ii) $*F$ is proportional to
$F^3$ provided the Lie algebra is such that $
E_a^2E_b$ is proportional to $E_b$.

We proceed now with the construction of
the $D=8$ octonionic instanton solution.
We note however that the conditions of Proposition 9 are too strong and we
can
obtain an interesting result by  considering a curvature 2-form where the
entries in each row come from different linear subspaces.

Suppose $F$ is an $so(8)$-valued gauge field 2-form.
We may  write $F= \sum_{i,j}\omega_{ij}E_{ij}$ where
$\{E_{ij}\}$ is  the
standard basis of skew-symmetric orthogonal
$8\times 8$ matrices.
The
$\{\omega_{ij}\}$
is a set of mutually orthogonal self-dual 2-forms.
The requirement that $F^2$ is self-dual in the Hodge sense
severely restricts the choice of $\omega_{ij}$'s. A possible choice
can be found as follows. Let
$\omega_{i8}$ be a basis for any linear subspace of self-dual 2-forms. Then
$\omega_{jk}$ for $j,k\ne 8$ are determined uniquely by the conditions
$(\omega_{jk},\omega_{i8})=0$ for $i=1,\dots 7$ and 
$\omega_{j8} \wedge \omega_{jk}=*(\omega_{j8} \wedge \omega_{jk})$ and
$\omega_{k8} \wedge \omega_{jk}=*(\omega_{k8} \wedge \omega_{jk})$.
Then by construction,
making use of the identities given above, we see that
the entries of the squared matrix
are 4-forms that are
self-dual in the Hodge sense.
Thus  $F$ would saturate the topological lower bound
$||F^2||^2 = \int_{M}tr(F \wedge F \wedge F \wedge F).$
It also turns out that $F^3=-90*F$.

All that remains to be done is to check
whether this $F$ could come from a potential $A$. To this end
we use the Bianchi identity $ AF - FA =-dF$.
It turns out that
$A$ can be determined in terms of the
derivative of a function  $\phi$. Substituting  into
the definition $F=dA-A^2$, we
obtain a system of second order differential equations  satisfied by
$\phi$. In these
equations all second  derivatives are determined and the
cofficients of the first
order derivatives
must be equal to each other.
Thus the solution is unique, and it can be shown that
in the Cartesian coordinate chart $\{x^{i}\}$ of $R^8$ we have
$$\phi=\left[\frac{1}{2}+x_1^2+x_2^2+x_3^2+x_4^2+x_5^2+x_6^2+x_7^2+x_8^2
\right]^{-2}.$$ This expression yields the octonionic instanton
solution [7].
It will be instructive to compare the above derivation of the
octonionic instanton solution
with the previous approaches based on $spin(8)$ matrices [16].
Let $F=\frac{1}{2}F_{ij} dx^{i} \wedge dx^{j}$. We may set
$F_{ij}= f \Sigma_{ij}$ where $f$ is a
function on $R^8$ to be determined
and  $8\times8$ matrices
$\Sigma_{ij}$'s form a basis for spin(8).
It can be shown that the non-linear self-duality
equation
$*F^2= F^2$ is satisfied as a consequence of
the algebraic properties of the spin matrices $\Sigma_{ij}.$
It must be clear from our exposition thus far that the two approaches are
equivalent.
\vskip 8mm
\noindent {\bf Acknowledgement}

\noindent T. Dereli wishes to thank Professor J. Lukierski and the organisers of the
{\bf IXth Max Born Symposium} for their hospitality.

\newpage
{\bf References}
\begin{description}
\item{[1]} J. M. Milnor, J. D. Stasheff, {\bf Characteristic Classes}
(Princeton U. P., 1974)

\item{[2]} F. R. Gantmacher, {\bf The Theory of Matrices} (Chelsea Publ.
Co., 1960)

\item{[3]} A. H. Bilge, T. Dereli, \c{S}. Ko\c{c}ak, Lett. Math. Phys.
{\bf 36}(1996)301

\item{[4]} A. H. Bilge, T. Dereli, \c{S}. Ko\c{c}ak,
{\em The geometry of self-dual 2-forms}, J. Math. Phys. (in print)
hep-th/9605060

\item{[5]} A. Trautman, Int. J. Theo. Phys. {\bf 16}(1977)561

\item{[6]} D. H. Tchrakian, Jour. Math. Phys. {\bf
21}(1980)166.

\item{[7]} B. Grossman, T. W. Kephart, J. D. Stasheff,
 Commun. Math. Phys.{\bf 96} (1984)431, Erratum: ibid,{\bf 100}(1985)311

\item{[8]} E. Corrigan, C. Devchand, D. B. Fairlie, J. Nuyts,
Nucl. Phys.{\bf B214}(1983)452

\item{[9]} F. A. Bais, P. Batenburg, Nucl. Phys. {\bf B269}(1986)363

\item{[10]} T. A. Ivanova, A. D. Popov, Theo. Math. Phys. {\bf 94}(1993)225

\item{[11]}M. Marcus and H. Minc, {\bf A Survey of Matrix Theory and
Matrix Inequalities} (Allyn and Bacon , 1964)

\item{[12]} I. M. Porteous, {\bf Clifford Algebras and The Classical Groups}
(Cambridge U. P.,1995)

\item{[13]} D. B. Fairlie, J. Nuyts, J. Phys.{\bf A17}(1984)2867

\item{[14]} S. Fubini, H. Nicolai, Phys. Lett.{\bf B155}(1985)369

\item{[15]} R. D\"{u}ndarer, F. G\"{u}rsey, C.-H. Tze, Nucl. Phys. {\bf B266}(1986)440

\item{[16]} G. Landi, Lett. Math. Phys. {\bf 11}(1986)171

\item{[17]} B. Grosmann, T. W. Kephardt, J. D. Stasheff, Phys. Lett. {\bf B220}(1989)431

\item{[18]} A. H. Bilge, {\em Self-duality in dimensions $2n>4$},   dg-ga/9604002

\end{description}
\end{document}